\documentstyle[amssymb,amstex,amscd,11pt]{article}

\def\ber{\begin{eqnarray*}}
\def\eer{\end{eqnarray*}}

\def\bc{\begin{center}}
\def\ec{\end{center}}
\def\be{\begin{equation}}
\def\ee{\end{equation}}
\def\ben{\begin{enumerate}}
\def\een{\end{enumerate}}
\def\bfg{\begin{figure}}
\def\efg{\end{figure}}
\def\bq{\begin{quote}}
\def\eq{\end{quote}}
\def\bd{\begin{description}}
\def\ed{\end{description}}

\def\p{\partial}
\def\w{\wedge}

\newcommand{\HX}{H^*(X)}
\newcommand{\HG}{H^*_G(X)}

\newcommand{\Hp}{H^*_G(pt)}
\newcommand{\Hpp}{H^*_{G'}(pt)}
\newcommand{\Hg}{H^*_{G'}(X)}

\newcommand{\pf}{\noindent{\em Proof.}\ }

\newcommand{\CC}{{\Bbb C}}

\newcommand{\ZZ}{{\Bbb Z}}
\newcommand{\QQ}{{\Bbb Q}}
\newcommand{\PP}{{\Bbb P}}

\newcommand{\gd}{\delta}

\newcommand{\gf}{\varphi}

\newcommand{\gl}{\lambda}

\newcommand{\gz}{\zeta}

\newcommand{\gS}{\Sigma}

\newcommand{\calm}{{\cal M}}

\newcommand{\calo}{{\cal O}}

\title{ Quantum
Cohomology of Partial Flag Manifolds and a Residue Formula
for Their Intersection Parings}

\author{Bumsig Kim \\ UC\, Berkeley}
\date{October 17, 1994}
\begin{document}

\maketitle
\begin{abstract} { As a generalization of our previous paper [GK],
 we formulate a residue formula and
some simple behaviors of equivariant quantum cohomology
applying to compute the quantum cohomology of
partial flag manifolds
  $F_{k_1,\cdots , k_l} $with a try to give a rigorous definition of
equivariant quantum cohomology.}
\end{abstract}
\section{Introduction}
Recently, the great effort of establishing Gromov-Witten invariants,
which arise in Mirror Symmetry of Calabi-Yau manifolds and give quantum
cohomology of compact semi-positive
symplectic manifolds, is succeeded by Ruan, Ruan-Tian,
McDuff-Salamon and Kontsevich [R][RT][MS][K]\footnote{Actually they did it
in different fields of manifolds.}.

But so far, a complete description of quantum
ring structure by a linear basis are known only for complex projective
space\footnote{ Also, some cases of lower dimensional manifolds are known.}.
For Grassmannian, there is an answer as a
family of algebra parameterized
by generators,
Chern classes of the tautological bundle
of ordinary cohomology, in papers [W][P] and in [ST] with
rigorous explanation of all details. In [GK], like the case of
Grassmannian, quantum cohomology of complete flag manifolds is  predicted
by relations of
Borel's generators
introducing equivariant quantum cohomology due to Givental or
equivariant Gromov-Witten invariants. The assumption therein to prove the
prediction is that
 the equivariant quantum
 cohomology of compact Kahler manifolds
is  well-defined, associative and a weighted-homogeneous
ordinary equivariant  $q$-deformation  of ordinary
cohomology.

\bigskip
The advantage of equivariant version of quantum cohomology is
its behaviors simply under product, restriction
and induction operation which we explain in section 4 (c.f. [GK]).

Using those properties under the assumption, equivariant quantum cohomology
is well-defined,
and stable maps due to Kontsevich [K],
we give a computation of
the equivariant quantum cohomology of
partial flag manifolds as a family of algebra parameterized by {\it
some generators}. Our hope is to prove or
check all details of a rigorous definition
of equivariant quantum cohomology elsewhere.
By some observations in residue properties and intersection parings
of complete flag manifolds, we derive a general theorem for
those of some manifolds which present their quantum cohomology as
regular functions on complete intersections, including partial flag manifolds.

\bigskip
Let $ F_{k_1,k_2,\cdots ,k_l} $ or simply $F$
denote the manifold of partial flags
\[  \CC ^{k_1} \subset \CC ^{k_1 + k_2}\subset \cdots \subset
 \CC^{k_1 + \cdots + k_l} \]
in $\CC ^n $ where  $ n=k_1 +\cdots + k_l \text { and integers } k_i > 0 $.
To describe the (equivariant) cohomology of it,
let us consider, for total Chern class
$c=1+c_1+...+c_k$ ($ c_i=$i-th Chern class),
$k\times k$ matrix $A(c)$ which is
\[\pmatrix
  c_1 & c_2 &...&...& c_k \\
  -1  & 0   &...&...& 0  \\
   0 & -1 &...&...& 0  \\
   \vdots & \ddots &\ddots & &0 \\
    0   & ...&...& -1& 0
\endpmatrix. \]
\noindent
Let $A(c^1,c^2,...,c^l)=$diag$(A(c^1),...,A(c^l))$
which is
\[\pmatrix
    A(c^1) & &0 \\
           &\ddots & \\
        0  &      & A(c^l)
\endpmatrix \]
where $c^i=1+c^i_1+...+c^i_{k_i}$ is the total Chern class of the
vector bundle with fiber $(\CC ^{k_i}/\CC ^{k_{i-1}})^*$
over  $F_{k_1,k_2,\cdots ,k_l} $
and $c_i$ is the i-th Chern class  of the universal $U(n)$-bundle over the
classifying space $BU(n)$.

  The equivariant cohomology algebra $H^*_{U(n)}(F_{k_1,k_2,\cdots ,k_l})$
 of flag manifolds $F_{k_1,...,k_l}$  as  $U(n)$-spaces
is canonically isomorphic to the quotient of the polynomial algebra
\[ \ZZ[c_1^{1},\cdots ,c_{k_1}^1,c_1^2,\cdots ,c_{k_2}^2,\cdots ,
c_1^{l},\cdots ,c_{k_l}^l,c_1,\cdots , c_n]\] by the ideal generated by
the  n coefficients of $f(\lambda )-g(\lambda )$
where $f$ and $g$ are the characteristic polynomials of
matrices $A(c^1,...,c^l)$ and $A(c)$ respectively.

For  quantum cohomology, consider nonnegative determinant
line bundles of fibers $(\CC ^{k_1 +\cdots +k_i})^*
\\ =Hom(\CC ^{k_1+\cdots +k_i},\CC) $ and the duals
$q_i$ in $H_{1,1}(F)\cap H_2(F,\QQ)$ of their first Chern classes.

\bigskip
{\bf Theorem I} {\em The equivariant quantum cohomology of partial
flag manifold $F_{k_1, \cdots ,k_l} , \text {where }
k_1  + \cdots + k_l =n$,
is canonically isomorphic to the quotient of the polynomial algebra
\[
\QQ [c_1^1,\cdots ,c_{k_1}^1,\cdots ,c^l_1\cdots ,c_{k_l}^l,
 q_1,\cdots ,q_{l-1}, c_1 ,\cdots ,c_l] \]
by the ideal generated by the coefficients of
\[ (\ast )\cdots\cdots
\text {det} (A(c^1,...,c^l) + C_{k_1,\cdots ,k_l}(q) + \gl )
 - \text {det} (A(c) + \gl )
\]
 where $C_{k_1,\cdots ,k_l}(q)$ is a matrix with $0$
entities except $(-1)^{k_i}q_i$ at
positions $(k_1 +\cdots + k_{i-1}+1,k_1+\cdots +k_{i+1})$
and $-1$ at positions $(k_1+...+k_i+1,k_1+...+k_i-1)$
for $i=1,\cdots ,l-1$ .}

\bigskip

Taking $c_i = 0$ in the
above relations, we get the quantum cohomology of partial
flag manifolds.

For the volume generating function\footnote{A correlation function
in topological $ \sigma $ model.}
of partial flag manifolds  which is  defined by $\Psi(\alpha )=
\sum_N\sum _d\frac {1}{N!}
q^d\Psi _d(\overbrace{\alpha ,...,\alpha}^{\text {$N$ times}} )$
for $\alpha \in H^*(X)$
and the notation $\Psi _d$\footnote{$\Psi _d=\widetilde\Phi _d
$ in [R]} in [MS],
we formulate a general theorem for some $G$-space $X$.

Ordinary cohomology classes will be called geometric classes.

\bigskip
{\bf Theorem II} {\em Suppose $X$ be a positive symplectic
manifold and $QH^*_G(X)$ is generated by geometric
even degree generators
$p_1,\cdots ,p_n$ with only $n$ algebraically independent
 relations$-$or regular sequence in each $q-$ $$\gS _1(p,q)-c_1,\cdots ,\gS _n
(p,q)-c_n$$
where $q$ are parameters from quantum cohomology and
$c_i, i=1,...,n$  are $G$-characteristic classes in $\Hp =\QQ [c_1,...,c_n]$.
Then the volume generating function is the  global residue integral
\[ \frac {a}{(2\pi \sqrt{-1})^n}\int \frac {exp(z,p)dp_1\w\cdots\w dp_n}{(\gS
_1(p,q)-c_1)\cdots
(\gS _n(p,q)-c_n)}
\]
for some suitable number $a$.
In particular, $a$=1 for $U(n)$-space partial flag manifolds with respect to
the relations in theorem I. Hence a partial answer of Gromov-Witten invariants
of genus zero is obtained for those generators of cohomology of partial flag
manifolds.}

\bigskip
When this paper was in preparation, we learned
that A. Astashkevich and V. Sadov
obtained the same result in computation
of quantum cohomology of partial flag manifolds [AS].

\bigskip

{\it Conventions.}
For the sake of simplicity, assume $H^*(X,\QQ)$ denoted by $H^*(X)$,
has even degree cohomology classes.
Hence it is convenient to count all dimensions and degrees
in complex units---one half of
those in ordinary (real) units throughout this paper except when we
say even degrees.

\bigskip
{\it Thanks.} I would like to express my sincere gratitude to my thesis
advisor Alexander Givental
for teaching me mirror symmetry phenomena and for encouraging  me to
produce this paper. I  would also like to  thank  Hung-wen
Chang for numerous helpful discussions.

\section {A Construction of Multiplications}
Let $R$ $ (R')$ be a commutative $\QQ$-algebra such that unity $1\in\QQ
\subset\QQ\cdot 1\subset R (R')$
and let $A$ be a finite dimensional $R$-free module
with a nondegenerate symmetric $R$-valued $R$-bilinear
pairing $<\cdot |\cdot >$
and unity $1\in\QQ\subset R\subset A$ and $a\in A$.
Suppose also $A$ has two basis $\{w_i\}$ and $\{ \tilde {w}_i\}$
such that $<w_i| \tilde {w}_j>=\gd_{i,j}$

In this set-up, we have the following useful observations.

\begin{description}
\item[i)]
Suppose $A$ is an $R$-algebra with $ra=ar$ for any $r\in R$
and $a\in A$. Then it has an $R$-multilinear
triple-pairing  $<\cdot |\cdot |\cdot >$ defined by
$$1)\cdots\cdots\cdots <a|b|c>=<ab|c>.$$
The invariant property  of $R$-valued product so that $A$ becomes a
Frobenius algebra\footnote{See [D] for definition.}
over $R$ is equivalent to a condition
$$2)\cdots\cdots\cdots \sum _i<a|b|w_i><\tilde {w}_i|c>
=\sum _i<a|w_i><b|c|\tilde {w}_i>.$$

The associativity of algebra $A$ is equivalent to condition
$$3)\cdots\cdots\cdots \sum _i<a|b|w_i><\tilde {w}_i|c|d>=
\sum _i<b|c|w_i><a|\tilde {w}_i|d>. $$

\item[ii)]
Suppose $A$ has a $R$-valued triple pairing satisfying  2) and 3), then
it becomes an unique Frobenius algebra over $R$ satisfying 1)

\item[iii)]
Let $f:R\rightarrow R'$ is $\QQ$-algebra homomorphism such that it
induces a Frobenius algebra over $R'$ from $A$ in ii).
Then $a\cdot _Rb=a\cdot _{R'}b$.

{\bf Example.}
Suppose a differentiable map from a compact oriented manifold  $X$
to another $Y$ induces
a $H^*(Y)$-free algebra $H^*(X)$
with two $H^*(Y)$-basis  $\{w_i\}$ and $\{ \tilde {w}_i\}$
such that $<w_i| \tilde {w}_j>=\gd_{i,j}$.
Then after a forgetting of $H^*(Y)$-module structure,
$H^*(Y)$-algebra $H^*(X)$ is $\QQ$-algebra $H^*(X)$.

\item[iv)]
If $f:A\rightarrow B$ be a $\QQ$-algebra homomorphism from $A$ to $B$,
sending $R$ to $R'$,
such that the induced map $A\otimes _RR'\rightarrow B$
is an $R'$-linear isomorphism,
then they are isomorphic as $R'$-algebras.

{\bf Example.}
{}From maps between compact oriented manifolds
$$\begin{CD}
X @<<< X' \\
@VVV @VVV \\
Y @<<< Y'
\end{CD}$$
we have maps between rings
$$\begin{CD}
A=H^*(X) @>>> B=H^*(X') \\
@VVV @VVV \\
R=H^*(Y) @>>> R'=H^*(Y')
\end{CD}$$

Then $H^*(X')$ as $H^*(Y')$-algebra
is isomorphic to $H^*(X)\otimes _{H^*(Y)}H^*(Y')$
if the condition in iv) is satisfied.

\item[v)]
Consider Frobenius algebras $A$ and $B$
with a $\QQ$ linear map $f:A\rightarrow B$.
If $R'f(A)=B$, $f(R)\subset R'$  and
$$\begin{CD}
A^{\otimes N}@> {f^{\otimes N}}>> B^{\otimes N} \\
@VVV                   @VVV \\
R @>>f> R'
\end{CD}
$$
is commutative for $N=2,3$ where vertical maps are pairings, then
$f$ is a $\QQ$-algebra homomorphism.

\pf
\ber
<f(a)f(b)|w>&=&<f(a)f(b)|\sum _ir'_if(c_i)>=\sum _ir'_i<f(a)|f(b)|f(c_i)> \\
&=&\sum _ir'_if(<a|b|c_i>)=\sum _ir'_i<f(ab)|f(c)>=<f(ab)|w>.
\Box
\eer

\end{description}
\section {Quantum Cohomology}

\bigskip
{\bf 3.1 Naive definition.}
Let $X$ be a compact connected Kahler manifold with the positive first Chern
class $c_1(TX)$ of the tangent bundle.
As an $\QQ$-free module, the quantum cohomology of X,
by definition, is the tensor product of ordinary cohomology
and $\QQ [q] $ where $q=(q_1,\cdots ,q_s)$ is a basis of
the space  $Q(X)$ generated by the closed Kahler cone of $X$, which is in
$H^{1,1}(X)\cap H^2(X,\ZZ)$.
But their ring structure is quite different from ordinary one $H^*(X)$
which can be defined, as in the context of the preceeding section,
by the intersections
of three cocycles $x, y$ and $w$ with  the Poincare pairing $ <\cdot,\cdot>=
\int _X\cdot\w\cdot : $

\begin{eqnarray*}    <x\cdot y,w> & = & < x,y,w> \text { by notation }  \\
                                  & = &  a\cap b\cap c \\
                                 & = & \int_X x\wedge y\wedge w
\end{eqnarray*}
where cycles $a, b$ and $c$ are denoted the Poincare dual of $x, y$ and $w$.

In more detail, for the diagonal class
$\sum _i \alpha _i\otimes \beta _i \in H^*(X\times X)=
H^*(X)\otimes H^*(X)$, which is
Poincare-dual to the homology class of the diagonal $X\subset X\times X$,
the product  is defined by
\[ x\cdot y = \sum _i <x,y,\alpha _i>\beta _i. \]

To give instanton corrections, let us think of the space $\cal M_d$  of
holomorphic maps from $\CC P^1$ to $X$
with given type $d=(d_1,\cdots ,d_n)\in H_2(X,\ZZ )\cap H_{1,1}(X)$
with respect to a coordinate $q_i$.
Then the ring structure of quantum cohomology  $QH^*(X,\QQ)$
 as a $\QQ [q] $-algebra is, by definition, given by a triple product
 \begin{eqnarray*}  <x*y,w>& = &<x|y|w> \text {by notation} \\
                    & = &\sum _{d}
                     \sum _{\gf \in \cal M_d \text { such that } \gf (0)\in a,
                           \gf (1)\in b \text { and } \gf (\infty )\in c}
                           \pm q^d  \\
                        & = &\sum _{d} q^d \int _{\cal M_d}
                 \gf ^*_0 (x)\wedge\gf ^*_1 (y)\wedge\gf ^*_\infty (w)
\end{eqnarray*}
where $\gf _0, \gf _1 \text { and }\gf_\infty $ are
 evaluation maps from $\cal M_d$ to $X$ at $0, 1 \text { and } \infty $.

{\bf 3.2 Remarks.}
The difficulties  of the rigorous definition  of quantum cohomology
 come from  when
the $\CC$-dimension of the moduli space $\cal M_d$ is $c(d)+\text {dim}X$ as
predicted
by the Riemann-Roch theorem where  $c(d)$ is the pull back of the first
Chern class of $X$ by a holomorphic map in $\cal M_d$, giving a nice
compactification of the moduli space $\cal M_d$ and showing  associativity
which was highly nontrivial and overcome by Ruan
and Tian for semi-positive symplectic manifolds using an
inhomogeneous term in Cauchy-Riemann equation [RT].
Recently also McDuff and Salamon [MS] give a complete definition by
Gromov compactification. There is an beautiful
algebro-geometric approach achieved by
Kontsevich and Manin [KM][M] for projective manifolds and analytic symplectic
manifolds using stable maps, stable curves and vertical fundamental classes.

\bigskip
Here we will follow the moduli space
defined by Kontsevich and Manin [KM] since, according to Kontsevich [K],
the moduli space of stable maps to convex manifolds---by definition, they
are manifolds such that
for any stable map f, $H^1(C,f^*(TX))=0$---are compact and
smooth as stacks\footnote{For example, homogeneous projective varieties
are convex.} and their motivic axiom seems easily applied
to equivariant version.
{\it So, throughout the paper, we consider
a convex manifold $X$ and the moduli
space $\overline {\cal M}_{0,3}(X,d)$ of stable maps in [M]
which is the collection of
all algebraic map from a connected compact reduced curve $C$ with genus zero,
marked three distinct points $x_1, x_2, x_3$ or symbolically $0, 1, \infty$
and at most ordinary double singular points
to $X$ such that every irreducible component of $C$ which
maps to a  point must have at least 3 special (i.e.\ marked or ordinary
double singular) points on its normalization.}

\bigskip
{\it Conventions.}
For shorthand, $\overline  {\calm}_d$ and $\check {a_i}$
will be denoted $\overline  {\calm}_{0,3}(X,d)$
and $\varphi _{x_i}(a_i), a_i\in H^*(X)$ respectively.
Note that $\overline {\cal {M}}_0= X$ so that
we get a \lq\lq $q$-deformation" ring structure.

{\bf 3.3 Volume generating functions.}
For even degree basis or generators $p_1,\cdots ,p_M$ of
ordinary cohomology ring, we can define the volume
generating functions $\Psi(z_1,...,z_M)$ introduced by Givental [GK];
$$ \Psi(z_1,\cdots ,z_M)= \int _Xexp (z_1p_1+\cdots +z_Mp_M) $$
where   exponentials are obtained from quantum multiplications in $X$.
Then, we have another equivalent definition of quantum cohomology of
$X$ by $\QQ [p,q]$ modulo an ideal generated by polynomials
$R(p,q)$ such that $ R(\frac{\partial }{\partial z},q)\Psi(z,q)=0 $.
We shall see why they are equivalent definitions in subsection 5.5.

{\bf 3.4 Grading.}
If we give a degree of $q$ the first Chern class of the pull back bundle
 over $\CC P^1 $ of the tangent bundle of $X$ by  a
 holomorphic map represented by $q$, then
$QH^*(X)$  will be a $\ZZ$-graded algebra.

\section {Equivariant quantum  cohomology}

{\bf 4.1  Definition.}
The classical equivariant cohomology $H^*_G(X)$  of a manifold  $X$
on which $G$ acts is
the ordinary cohomology of the homotopy quotient $X_G$$-$or $X^G$ later for
partial flag manifolds$-$of $X$ and $EG$,
 i.e. $X\times _G EG$
as a $H^*(BG)$-module.
In greater detail, consider the associated fiber  bundle
$  X_G @>>{\pi}> BG $ with fiber $X$
from a fixed  universal principal $G$-bundle $EG\rightarrow BG$.
Then the ordinary cohomology of $X_G$ as a $H^*(BG) =H_G^*(pt)$-module
by the pull back of
the projection $\pi$ is called the equivariant cohomology of $G$-space
$X$.
Suppose $H^*_G(X)$ is a  $H^*(BG)$-free module
canonically isomorphic to $\HX\otimes\Hp$,
then we may consider the ordinary
equivariant cohomology as a Frobenius algebra
over coefficients $H_G^*(pt)$ provided with equivariant $N$-intersection
$\Hp$-indices,
namely for $p_1,...,p_n\in H^*(X_G)$---$H^*(BG)$-module---
and a cycle $C$ in $BG$,
$<p_1,\cdots ,p_N>[C]=(p_1\cdots p_N)[\pi ^{-1}(C)]$ which is
$H^*_G(pt)$-multi-linear\footnote{I have learned this
method only from Givental.}.

Recall some facts on equivariant cohomology to fully understand our
definition [AB][G][GK].
\begin{description}
\item[i)]
 If $G$-action on $X$ is Hamiltonian for a symplectic form or a
 Kahler form,  then  $H^*_G(X)$ is a $H^*(BG)$-free module
isomorphic to $H^*_G(pt)\otimes H^*(X)$ so that
$H^*_{G'}(X)$ is canonically isomorphic to
 $H^*_{G'}(pt) \otimes _{H^*_G (pt)} H^*_G(X)$ for
a Lie subgroup $G'\subset G$.
\item[ii)]
An $G$-action from an algebraic $G_\CC$-action
on a complex projective
manifold is always Hamiltonian.
\item[iii)]
$$\begin{CD}
\HG @>>{\cong }> \Hp\otimes\HX \\
@V{\pi _!} VV @VV{\text {integration over $X$}}V\\
\Hp@>>{\cong }>\Hp\otimes H^*(pt)
\end{CD}
$$
is commutative where $\pi _!$ is denoted
the push forward of $\pi :X_G\rightarrow
BG$.
\item[iv)]
The pairing
$$\HG\otimes\HG@>>{\text {product}}>\HG
@>>{\text {push forward}}>\Hp$$ is nondegenerate and its
associated determinant
is in scalar $\QQ$.
\end{description}
Suppose $X$ is a compact convex Kahler manifold
provided with a holomorphic action of compact Lie group $G\subset G_\CC $.
The additive structure of equivariant quantum cohomology
$QH^*_G(X)$ of $G$-space
$X$ is $H^*_G(X)\otimes \QQ [q]$
where $q$ is a basis of $Q(X)$.

Then the quantum ring structure of $QH^*_G(X)$
 as $H^*_G(pt)\otimes \QQ [q]$-module is, by definition [GK],
given by triple products of three cycles $a, b$ and $c$
 and their Poincare-duals $x, y$ and $w$;
\begin{eqnarray*}  <x*_{X_G}y,c>[C]  &=& < x| y| w>[C] \ \ \text { by notation
} \\
                & = &\sum _{\overline {\cal M}_d} \sum _{ \gf \in \cal M_d[C]
                       \text { such that } \gf (0)\in a,
                  \gf (1)\in b \text { and } \gf (\infty ) \in c } \pm q^d   \\
                & = & \sum _{d}q^d \int _{\overline {\cal M}_d[C]}
                        \gf ^*_0 (x))\w\gf ^*_1(y) \w\gf _\infty ^* (w)
\end{eqnarray*}
where $a$, $b$ and $c$ are  finite co-dimension
 cycles in $X_G$ and $\overline {\cal M}_d[C] $ is the space of all stable maps
from from stable curves of genus zero marked three points $0, 1, \infty$
to $X_G$ of type $d$ such that their image under the projection $\pi$
are points at the  finite dimensional  cycle $C$ in $BG$---let us
call them vertical stable maps. Notice that $H^{1,1}(X)\cap
H^2(X,\ZZ )$ is a sublattice $\subset H^2(X_G)$ so that we can canonically
count types of vertical stable maps by $d$.

\bigskip
The diagrams
\ber
& & \overline {\cal {M}}_d(X)_G=_{\text {claim}}\overline {\cal {M}}_d(X_G)
@>>{\pi ^d}>
BG \\
& & H^*(\overline {\cal {M}}_d(X)_G)^{\otimes 3}
@>> {\pi _!^d} >
\Hp
\eer
should give an almost complete picture for our definition,
where $\overline {\cal {M}}_d(X)_G $ is the homotopic quotient
of the moduli space $\overline {\cal {M}}_d$
of genus zero stable maps of  genus zero and with three
marked points $0,1,\infty$ to $X$. {\it Here we assume that the above
picture gives an associative equivariant (quantum cohomology) ring
structure.}

\bigskip

In [GK] simple behaviors of volume generating functions formulated.
But here
we would like to point out the simple behaviors
of equivariant quantum cohomology:

{\bf 4.2.} It is a $H^*_G(pt)\otimes \QQ [q] \subset QH^*_G(X)$-module.

{\bf 4.3 Product.}  Let $X'$ and $ X''$ be compact Kahler $G'$-and $G''$-spaces
 respectively, then
the quantum cohomology of $G'\times G''$-space $X'\times X''$ is the tensor
product of those of $X'$ and $X''$ since $H^*_{G'}(pt)\otimes H^*_{G''}(pt)$ is
 canonically $H^*_{G'\times G''}(pt)$.

{\bf 4.4 Restriction\footnote{One might convince oneself by
the functorial nature of equivariant version.}.}
Let X be a compact Kahler $G$-space
and $G'\subset G$ be a
Lie subgroup. Considering X as a $G'$-space, we obtain an $ X$-bundle
$X_{G'} \rightarrow BG'$ induced, as a bundle, from $X_G\rightarrow
 BG$ by means of the natural map $\pi :BG'\rightarrow BG$ of classifying
spaces and corresponding map of total spaces $\gz  : X_{G'}\rightarrow
 X_G$ with the fiber $G/G' $.
In this setup we have the following lemma:

\bigskip
{\bf Lemma }{\em
$QH^*_{G'}(X)$ is canonically isomorphic to $H^*_{G'}(pt)\otimes _{
H^*_G(pt)} QH^*_G(X)$ as $\Hpp$-algebras.}

\bigskip
{\em Proof}.
It is enough to show that the induced $\QQ$-linear map $\gz ^*$
is a quantum ring homomorphism by iv) in section 2 and
i) in 4.1.
 From v) in section 2, we only need to prove that
$$
\begin{CD}
Q\Hg ^{\otimes 3} @<{\gz ^{*\otimes 3}}<< Q\HG ^{\otimes 3} \\
@V{3\text{-intersection index}}VV @VV{3\text{-intersection index}}V \\
\Hpp\otimes\QQ [q] @<{\gz ^*}<< \Hp\otimes\QQ [q]\\
\end{CD}
$$
is commutative.
But it follows because for $a_1,a_2,a_3\in\HG$ and corresponding forms
$\check {a_1},\check {a_2},\check {a_3}\in H^*_G(\overline {\cal {M}}_d)
=H^*(\cal {M}_dG)$,
\ber
\pi _!(\gz ^*a_1\gz ^*a_2\gz ^*a_3) &=& \sum _d q^d
\pi _!^d(\check {\gz ^*}a_1\check {\gz ^*}a_2\check {\gz ^*}a_3) \\
&=&\sum _d q^d\pi _i^d(\gz ^*(\check{a_1}\check{a_2}\check {a_3}) \\
&=&\sum _dq^d\gz ^*p_!^d(\check {a_1}\check {a_2}\check {a_3}) \\
&=&\gz ^*p_!(a_1a_2a_3)
\eer
where $\pi _!^d$ and $p_!^d$ are push forwards as in a commutative diagram
$$
\begin{CD}
H^*(\overline {\cal {M}}_d(X)_{G'}) @<<{\gz ^*}<   H^*(\overline {\cal
{M}}_d(X)_G) \\
@V{\pi ^d _!}V{\text {fiber \/}\overline {\cal {M}}_d}V @V{p ^d_!}V{\text
{fiber \/} \overline {\cal {M}}_d}V \\
\Hpp @<<{\gz ^*}<\Hp \ .
\end{CD}
$$
$\Box$

{\bf 4.5 Induction}. Let $G'\subset G$ be a subgroup with a simply connected
compact Kahler quotient $G/G'$ and
$Y$ be a compact Kahler $G'$-space. We construct a compact Kahler $G$-space
$X=G\times_{G'} Y$ and call it {\em induced} from $Y$ (like induced
 representation). In fact $X$ is fibered over $G/G'$ with the fiber $Y$.
The homotopic quotient  spaces of $X$ and $Y$ coincide.
Let $p''$ be a basis of nonnegative classes in $H^2(G/G')$ lifted
to $X$ and $p=(p',p'')$ be its extension to such a basis in $H^2(X,\ZZ)\cap
H^{1,1}(X)$. By a Kahler volume,
we find that the curves vertical in the bundle $X\rightarrow G/G'$
have $d''=0$ and {\it vice versa} where $d''$ is a degree respect to $p''$
so that $\overline {\cal {M}}_{d',d''=0}(X)_G
=\\overline {cal {M}}_{d'}(Y)_{G'}$.
Via $BG'\rightarrow BG$,
$\HG$ is $H^*_{G'}(Y)$ as
$H^*_{G}(pt)$-module and they have a nice relation also in quantum algebra,
namely

\bigskip
{\bf Lemma }
{ $QH^*_{G}(X)|_{q''=0}$ is  $QH^*_{G'}(Y)$ as $\Hp$-algebras
where $q'$ and $q''$ are denoted duals of $p'$ and $p''$ respectively.}

\bigskip
{\em Proof\/\footnote{See the preceeding footnote.}.}
First, notice that
$$\begin{CD}
\overline {\cal {M}}_{d'}(X)_G @= \overline {\cal {M}}_{d'}(Y)_{G'} \\
@VVV @VVV \\
BG @<<< BG'
\end{CD}$$ is commutative.
Using iii) in section 2 if the diagram
$$
\begin{CD}
QH^*_{G'}(Y)^{\otimes 3}@=QH^*_G(X)^{\otimes 3}\\
@VV{p_!}V @VV{\pi_!}V \\
\Hpp\otimes\QQ [q'] @>> {f_!}> \Hp\otimes\QQ [q']
\end{CD}
$$
is commutative, the lemma follows.
But
\ber
\pi _!(a_1a_2a_3) &=&\sum _dq^d\pi _!^d(\check {a_1}\check {a_2}\check {a_3})
=\sum _dq^d f_!p_!^d(\check {a_1}\check {a_2}\check {a_3}) \\
&=&f_!p_!(a_1a_2a_3)
\eer
since $f_!p_!=\pi _!$ by a functorial property of push forwards.
Thus, the above diagram is commutative.
$\Box$
\bigskip

{\bf 4.6 Grading}.
As the quantum cohomology, our equivariant version, also,
has a natural $\ZZ$-gradation by
letting deg$q=c(d)$ and degree of cohomology
classes = ordinary degree, which will
play a crucial role on our computation in next section.

\section{ Computation  }
In this section $G=U(n)$, $X=F_{k_1,\cdots ,k_l}$,
$n=k_1+...+k_l$ and $k=k_1+...+k_i$.

\bigskip
{\bf 5.1.}
First, let us specify a basis $ \{q_i\}$ of $H_{1,1}$ of the flag
 manifold by the duals of
first Chern classes of the duals   of the  determinant line bundles
with fibers $\CC ^{k_1+ \cdots +k_i}$.
Since partial flag manifolds  $F_{k_1,\cdots ,k_l}$ are  $U(n)$-
Kahler manifolds, $F_{k_1, \cdots ,k_l}^{U(n)}=
F_{k_1,\cdots ,k_i}^{U(k)}\times F_{k_{i+1},\cdots ,k_l}^{U(n-k)}$,
integral homology of $H_{1,1}(F_{k_1,
\cdots ,k_i})=\ZZ q_1\oplus\cdots\oplus\ZZ q_{i-1},$
integral homology of $ H_{1,1}(F_{k_{i+1},\cdots ,k_l})=
\ZZ q_{i+1}\oplus\cdots\oplus\ZZ q_{l-1}$
and they have the simple behavior of the
equivariant cohomology, i.e. they are $H^*_{U(n)}(pt,\ZZ )$-free
module canonically isomorphic to $H^*_{U(n)}(pt)\times H^*(X)$,
we can state

\bigskip
{\bf Proposition}
1. Restriction to identity group.
\begin{eqnarray*}
QH^*(F_{k_1,\cdots ,k_l})&\cong &QH^*_{id}(F_{k_1,\cdots ,k_l})
\\ &\cong &
QH^*_{U(n)}(F_{k_1,\cdots ,k_l})/<H^+_G(pt)>
\end{eqnarray*}
{\em where $id$ is the trivial group and $<H^+_G(pt)>$ is the
ideal generated by positive degree elements.}

2.  Product. \begin{eqnarray*}  & & \mbox{}
QH^*_{U(k)\times U(n-k)}(F_{k_1,\cdots ,k_i}
\times F_{k_{i+1},\cdots ,k_l})  \\
&\cong & QH^*_{U(k)}(F_{k_1,\cdots ,k_i})\otimes _\QQ
QH^*_{U(n-k)}(F_{k_{i+1},\cdots ,
k_l}).
\end{eqnarray*}
3. Induction.
{\em As $H^*_{U(n)}(pt)$-algebras,
\begin{eqnarray*}
  & & \mbox{}  QH^*_{U(n)}(F_{k_1,\cdots ,k_l})\otimes _{\QQ [q]}
\QQ [q_1,\cdots,\hat{q} _i,\cdots ,q_l] \\
&\cong & QH^*_{U(k)\times U(n-k)}(F_{k_1,\cdots ,k_i}\times F_{k_{i+1},\cdots ,
k_l})
\end{eqnarray*}
where $q_i$ is dropped.}

\bigskip
{\bf 5.2 Theorem.}([W][P][ST])
 {\em  For Grassmannian $ G(n,k)=F_{k,n-k}$ the theorem
 in the introduction holds.}

\bigskip
{\em Proof. }  Since ordinary equivariant cohomology  has relations
det$A_{k,n-k}(c^1,c^2)$=det$A_n(c)$
and we have quantum correction only on top degree,
namely $c^1_kc^2_{n-k}=(-1)^kq$ by Witten [W], the proof follows.
$\Box$

\bigskip
{\bf 5.3 Lemma.}(c.f. [GK])
 {\em  For $ l>2$, suppose that a quasi-homogeneous
 relation of the form \begin{eqnarray*}  & &
 (\gl ^{k_1} + c_1^{1}\gl ^{k_1-1}
 + ... + c_{k_1}^1 ) \cdots (\gl ^{k_l}
 + c_1^l\gl ^{k_l-1} + ... + c_{k_l}^l)
 - (\gl ^{n} + c_1\gl ^{n-1} + \cdots
 + c_n) \\
& = &O(q_1 ,\cdots ,q_{l-1}) [ \gl ,q,c^i,c ]
\end{eqnarray*}
is satisfied in the quantum equivariant cohomology algebra of  partial flag
manifolds  $F_{k_1,\cdots ,k_l}$ modulo $q_i$ for each  $i=1\cdots l-1$.
Then this relation holds identically ( i.e. for all $q$).}

\bigskip
{\em Proof. } Since each line bundles $\bigwedge ^*\CC ^{k_1+\cdots +k_i}$
over the flag manifold is the induced bundle
of the dual of the determinant of universal bundle over Grassmannian $G(n,k)$,
we have deg $q_i = k_i + k_{i+1}$.
Hence LHS-RHS of the
above equation is divisible by $q_1\cdots q_{l-1}$ of degree $> n$ which
completes the proof.
$\Box $
\bigskip

{\bf 5.4 Proof of relations in theorem I.}
We shall use induction on $l$ in $F_{k_1,\cdots ,k_l}$.
When $l$=1, it is the case of Grassmannian which has shown in  theorem 5.2.
When $l>1$ it suffices to show by lemma 5.3 that
 the relation ($\ast$) in theorem I  from the introduction
holds modulo each $q_i$ in order to get the relation.
But this follows from statements 2 and 3 in the proposition and induction
hypothesis.
The statement 1 in proposition yields relations
of the non-equivariant version of theorem I.

{\bf 5.5 No more relations.}

Using the induction on degree, one can prove that
$\{ c_1^1,...,c_{k_1}^1,...,c_1^l,...,c_{k_l}^l\}$
generate the equivariant quantum cohomology and there are no more relations
counting ranks over $\QQ $ (see [BT][ST] for detail).
One may use a formula for Poincare series to count ranks over $\QQ $ explicitly
which depend only on degrees of generators and polynomial relations
(see [BT] for detail).
Now we are done.

\section {Residue formulas  for volume generating functions}
{\bf 6.1 Proof of theorem II for non-equivariant case.}
Let $X$ be a compact positive Kahler manifold with $H^*(X)$ generated by
even degree elements $x_1,\cdots ,x_n$.
Suppose $\CC [x,q]$ modulo weighted homogeneous polynomials
$f_1(x,q),\cdots ,f_n(x,q)$  presents the
quantum cohomology of some manifold  where
deg$_\CC(x_i)=d_i$ which is the half of the usual one and $ f={f(x,q),
\cdots ,f_n(x,q)}$ is a regular sequence (or equivalently
defines a complete intersection)
for each parameter $q$---for each complex value of $q$---and
$x=(x_1,\cdots ,x_n)$ $ q=(q_1,\cdots ,q_s)$.
Then we have  the following properties of residues and regular sequences in
graded rings [BT][GH][T].
\begin{itemize}
\item Global Duality says the pairing $(\gf,\phi )=\text {Res}_f(\gf\phi )
=\int \frac {\gf\phi dx_1\cdots dx_n}{f_1...f_n}$ is
 nondegenerate on the ring $\CC [x]/<f> $,
where
$<f>$ is the ideal
generated by $f_1,...,f_n$.
\item Global Residue Theorem on
weighted projective space $\PP ^n_{1,d_1,d_2,\cdots ,d_n}$, which is
a variety so that the singular set is of co-dimension at least 2 and on which
Stokes' Theorem can be
applied, implies
Res$_f(g)=0$ if deg$g\le$ deg$f-\sum _i d_i -1$, deg$f=\sum \text {deg}f_i$.
\item
Res$_f(g)=0 $ if deg$g\ge$ dim$X$
since $f$ is come from the cohomology of a manifold.
\item dim$_\CC X=$ deg$f-\sum _id_i$
since $f$ is a regular sequence.

\item Residues algebraically depend on parameters $q$.
\item Since Res($\frac {df}{f})$ are integers,
 Res($\frac {df}{f})$ = Res($\frac {df(x,0)}{f(x,0)})=$
 local Res($\frac {df(x,0)}{f(x,0)})$ at zero which is \\
 dim$_\CC \calo /f(x,0)= $dim$_\CC \CC[x]/f(x,0)=$Euler number of $X$
where $\calo $ is the local  ring of holomorphic functions at $0\in\CC ^n$.
\item Computing the rank of a graded ring   by a method in [BT],
$$\frac {\prod _i \text {deg}f_i}{\prod _i\text {deg}x_i}
=\text {dim}_\CC\frac { \CC[x]}{ <f(x,0)>}$$ which is the Euler number of $X$.
\end{itemize}
Thus, we conclude that the evaluation map $<\cdot >$,
 which is the integration over $X$,
on the quantum cohomology of $X$ is nothing but
the residue map with respect to $f$ up to a constant multiplication
$-$note that $<|\frac{\partial f(x,q)}{\partial x}|>$ does not depend on $q$ by
degree counting where $|\frac {\partial f}{\partial x}|$
is the determinant of the Jacobian of $f=(f_1,
...,f_n)$ with respect to $x=(x_1,...,x_n)$.

{\bf 6.2 Computing the constant term.}
First, notice that $\gS _1,...,\gS _n$ are algebraically independ- \\
ent---which
is a result of a well-known fact that the
elementary symmetric polynomials are algebraically independent---
so that they form a regular sequence for each $q$.
To compute the constant for partial flag manifolds and $\gS _1,\cdots ,
\gS _n$, i.e. to prove Theorem II for non-equivariant cases of
partial flag manifolds, let us consider the following obvious fiber bundles
and the corresponding integrations along fibers
$$\left[
\begin{CD}
F(n) @>{F_{k_1}\times\cdots\times F_{k_l}}>> F_{k_1,\cdots ,k_l} \\
@VVV                          @VVV \\
\text {pt}@=                \text {pt}
\end{CD}
\right]\left[ \begin{CD}
H^*(F(n)) @>\pi _* >> F_{k_1,\cdots ,k_l}\\
@V{<\cdot >'}VV        @VV{<\cdot >}V \\
\CC  @=                    \CC
\end{CD}
\right]
$$

where $F(n)=F_{1,...,1}$.
Let use $u=(u_1,...,u_n)$ instead of $c^1_1,...,c^n_1$'s
for complete flag manifolds.
 We know $n!=<|\frac {\partial f}{\partial c}|>'$
by the residue formula of volume generating functions at [GK].
$n!=<|\frac {\partial f}{\partial u}|>'
=<|\frac {\partial f}{\partial c}|>\pi _*|\frac {\partial c}{\partial u}|
=k_1!\cdots k_l! <|\frac {\partial f}{\partial c}|> $;
consequently, $<|\frac {\partial f}{\partial c}|>=
$Euler number of $F_{k_1,\cdots
,k_l}$ and Theorem II for non-equivariant cases.

{\bf 6.3 Proof of theorem II.}
Let $V$ be the global residue
\[ \frac {a}{(2\pi \sqrt{-1})^n}\int \frac {exp(z,p)dp_1\w\cdots\w dp_n}{(\gS
_1(p,q)-c_1)\cdots
(\gS _n(p,q)-c_n)}
\]
and let $\Psi$ be the volume generating function of $X$ with respect to
even degree generators $p_1,...,p_n$. Then $\Psi =V$ if $c_1=...=c_n=0$
by subsection 6.1.
But since $(\gS_i(\frac{\p}{\p z},q)-c_i)V(Psi )=0$ for $i=1,...,n$
and $ V$ is holomorphic on variable $c_i$ by a well-known property---trace
formula---of residues,
$V$ is determined by leading terms in expansion with respect to
$c_1,...,c_n$
and hence equal to $\Psi$.

\bigskip
{\it e-mail address}: bumsig@@math.berkeley.edu
\end{document}